# Overlapping community detection in signed networks

Y Chen[1], B Z Tang[1], X L Wang[1,2] and B Yuan[1]

[1] Department of Computer Science and Technology, Shenzhen Graduate School, Harbin Institute of Technology, Shenzhen 518055, China

[2] School of Computer Science and Technology, Harbin Institute of Technology, Harbin 150001, China

E-mail: chenyi@hitsz.edu.cn, tangbuzhou@gmail.com, wangxl@insun.hit.edu.cn and yuanbo@hitsz.edu.cn

**Abstract**

Complex networks considering both positive and negative links have gained considerable attention during the past several years. Community detection is one of the main challenges for complex network analysis. Most of the existing algorithms for community detection in a signed network aim at providing a hard-partition of the network where any node should belong to a community or not. However, they cannot detect overlapping communities where a node is allowed to belong to multiple communities. The overlapping communities widely exist in many real world networks. In this paper, we propose a signed probabilistic mixture (SPM) model for overlapping community detection in signed networks. Compared with the existing models, the advantages of our methodology are (i) providing soft-partition solutions for signed networks; (ii) providing soft-memberships of nodes. Experiments on a number of signed networks show that our SPM model: (i) can identify assortative structures or disassortative structures as the same as other state-of-the-art models; (ii) can detect overlapping communities; (iii) outperform other state-of-the-art models at shedding light on the community detection in synthetic signed networks.

**Keywords**: Analysis of algorithms; Statistical inference; Networks; Random graphs.

**PACS**: 89.75.Hc, 89.65.-s

## 1. Introduction

Complex networks [1] provide a powerful tool for representing many real world complex systems, such as information systems [2, 3], social systems [4, 5], ecological systems [6], and others [1, 7, 8]. The task of complex network analysis is to identify the network's properties including network structure. Newman firstly introduced two types of network structures: assortative structure and disassortative structure [9]. The assortative structure—also called community structure—is a type of network structures in which most edges are within a group. The disassortative structure is a type of network structures in which most edges are across groups. For these two types of network structures, a large number of effective techniques have been proposed during the last several years, such as Potts model [10] and modularity model [11]. A detailed survey about them was presented by Fortunato [12].

Networks considering both positive and negative links are called signed networks. The community structure in signed networks is different from the assortative/disassortative structure in un-signed networks. In signed networks, most edges within a community are positive links, and most edges across communities are negative links [13]. The signed networks have gained considerable attention from different scientific disciplines, i.e. biology [14], computer [15], and social sciences [16]. For example, in a social network, positive links may denote friendship, agreement or trust whereas negative links may denote hostility, disagreement or distrust. Many studies have been presented for community detection in signed networks. Yang et al. proposed an agent-based approach to extract community structures by performing a random walk on positive links [17]. Gomez et al. extended the modularity method from un-signed networks to signed networks for community detection [18]. Traag proposed an algorithm based on the Potts model to find community structures in signed networks with only negative links [19]. Shen et al. provided a statistical probability model based on the mixture model to detect the disassortative structures of signed networks with both positive and negative links [20].

Most of the existing algorithms for community detection in a signed network aim at providing a hard-partition of the network where any node should belong to a community or not. However, they cannot detect overlapping communities where a node is allowed to belong to multiple communities. The overlapping communities widely exist in many real world networks. For example, a person in social networks may belong to both family and hobby groups. In signed networks, a node is overlapping on the condition that it connects with nodes in other communities by positive links or connects with nodes in the same community by negative links, i.e. the node E and node F shown in figure 1. Recently, several methods have been proposed for overlapping community detection in networks with only positive links, which may fall into three categories: Clique Percolation Method (CPM) [21], fuzzy clustering-based method [22-24] and mixture models [25-29]. The CPM supposes that edges within a community are likely to form a clique due to their high density whereas edges across communities are unlikely to form a clique. The fuzzy clustering-based method uses fuzzy relation to describe the case of a node belonging to more than one community [22]. Dunn used a fuzzy probability to describe a node belonging to a community (i.e., $c$-means clustering) [23]. Zhang et al [24] provided a method to approximately map nodes into a dimensional space by combining a modularity function, spectral mapping and fuzzing clustering. The mixture models model a network using generative graphical models. There are two types of mixture models for network community detection. The first are stochastic block-based models [26, 28], which generate a network from a perspective on node. In a stochastic block model, each node is assigned to a block, group, or community. And undirected edges then are placed independently between node pairs with probabilities, from a function of the group memberships of the nodes. The second are probabilistic mixture-based models [25, 27, 29], which generate a network from a perspective on edge. The probabilistic mixture model is inspired by the probabilistic latent semantic analysis [30] for text mining, and introduced for community detection in the Ref. [29]. Instead of assigning each node to a specific community, a probabilistic mixture model assigns each edge to one of blocks, groups, or communities with a probability, and then picks up nodes of the edge from the corresponding blocks. The latent variables in stochastic block models operate on vertices, while that in probabilistic mixture models operate on edges [31]. Both types of mixture models are suitable for network community detection. The main difference between them is that the stochastic block models can be used not only for network community detection, but also for link prediction [32, 33] from the node perspective; but the probabilistic mixture models can only be used for network community detection from the edge perspective. For community detection, stochastic block models usually perform well on synthetic networks, but poorly on many real-world networks [34], whereas the probabilistic mixture models perform well on both synthetic and real-world networks.

Although several methods have been proposed for overlapping community detection, most of them are limited to networks with only positive networks. They do not work in signed networks. In this paper, we proposed a novel probabilistic mixture model based on expectation-maximization (EM) method [35], called signed probabilistic mixture (SPM) model, to detect overlapping communities in undirected signed networks. It is a variant of the probabilistic mixture model which generates positive and negative links with different probabilities. To give a clear description, we presented an illustrative undirected signed network as shown in figure 1. For the signed network, our model will provide a correct overlapping partition, i.e. the community (A, B, C, D, E, F) and community (E, F, G, H, I). The advantages of our method are (i) providing soft-partition solutions in signed networks, such as nodes E and F belonging to two communities simultaneously; (ii) providing soft-memberships, which quantify "how strongly" a node belongs to a community. Experiments on a number of signed networks show that our SPM model (i) can identify assortative structures or disassortative structures as the same as other state-of-the-art models; (ii) can detect overlapping communities; (iii) outperform other state-of-the-art models at shedding light on the community detection in synthetic signed networks.

The remaining part of this paper is organized as follows. Section 2 presents the SPM model. Section 3 discusses the performance of the SPM model on the signed network with only positive links or negative links. Experiments are presented in section 4. Section 5 draws conclusions.

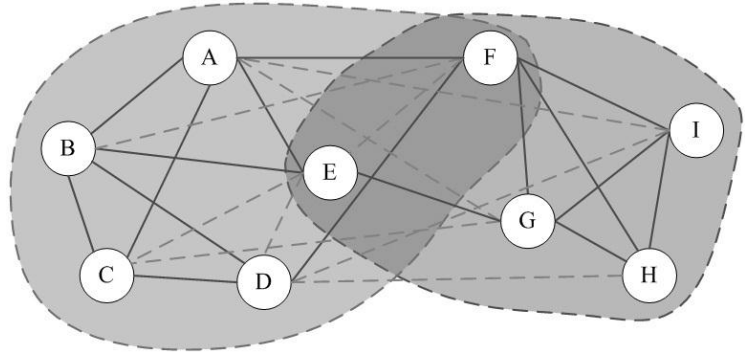

**Figure 1.** An undirected signed network with two overlapping communities (A, B, C, D, E, F) and (E, F, G, H, I). The solid lines denote positive links and the dotted lines denote negative links.

## 2. The signed probabilistic mixture model (SPM)

Before introducing the signed probabilistic mixture model, we give a brief definition of an undirected signed network. Generally, a network is represented by an adjacency matrix $A$ with $n$ dimensions. We use $E$ to denote the edge set, $A_{ij}(i \neq j)$ to denote the edge between node $i$ and node $j$. In addition, $E^+$ and $E^-$ are used to denote the positive and negative links in a signed network, respectively. It is easy to understand that: $E_{ij}^+ = A_{ij}$ if $A_{ij} > 0$; $E_{ij}^- = -A_{ij}$ if $A_{ij} < 0$. If there is no edge between node $i$ and node $j$, $E_{ij}^+ = E_{ij}^- = A_{ij} = 0$.

In the SPM model, given $K$ communities, an edge can chose a community pair from $K \times K$ possible community pairs because each node of the edge can chose a community from $K$ possible communities. (Here, $K$ is a predefined number of communities in a network). Formally, we use $\omega_{rs}$ to denote the probability of a edge choosing a community pair $\{r,s\}(1 \leq r,s \leq K)$, normalized by the constraint $\sum_{rs} \omega_{rs} = 1$. Specially, $\omega_{rr}$ is the probability of an edge locating in community $r$ (i.e., a positive link in community $r$); $\omega_{rs(r \neq s)}$ is the probability of an edge locating between community $r$ and community $s$ (i.e., a negative link between community $r$ and community $s$).

When an edge chooses a community pair $\{r',s'\}$, the SPM model actually determines a node pair $\{i,j\}$ from the corresponding community pair $\{r',s'\}$ with a probability. We use $\theta_{ri}$ to denote the probability of a community $r$ selecting node $i$. For all communities selecting node $i$, $\sum_i \theta_{ri} = 1$. It should be noted that the SPM model assumes every community contains every node with a probability. Similarly, $\theta_{sj}$ denotes the probability of community $s$ selecting node $j$.

In a signed network, an edge $E_{ij}$ (either $E_{ij}^+$ or $E_{ij}^-$) is generated as follows:

1) Check $E_{ij}$ belongs to $E^+$ or $E^-$. If $E_{ij}$ belongs to $E^+$, follow steps 2) 3) 4), otherwise follow steps 5) 6) 7);
2) Select a community $r'=r$ for a positive edge with probability $\omega_{rr}$;
3) Select node $i$ from the community $r'$ with probability $\theta_{ri}$;
4) Select node $j$ from the community $r'$ with probability $\theta_{rj}$;

5) Select two different communities $r'=r$ and $s'=s$ for a negative edge with probability $\omega_{rs(r \neq s)}$;
6) Select node $i$ from the community $r'$ with probability $\theta_{ri}$;
7) Select node $j$ from the community $s'$ with probability $\theta_{sj}$.

Overall, the probability of a positive edge $E_{ij}^+$ can be written as

$$P(E_{ij}^+ | \omega, \theta) = \sum_{rr} \omega_{rr} \theta_{ri} \theta_{rj} \tag{1}$$

And the probability of a negative edge $E_{ij}^-$ can be written as

$$P(E_{ij}^- | \omega, \theta) = \sum_{rs(r \neq s)} \omega_{rs} \theta_{ri} \theta_{sj} \tag{2}$$

We unify them into the probability of an edge $E_{ij}$:

$$P(E_{ij} | \omega, \theta) = (\sum_{rr} \omega_{rr} \theta_{ri} \theta_{rj})^{E_{ij}^+} \cdot (\sum_{rs(r \neq s)} \omega_{rs} \theta_{ri} \theta_{sj})^{E_{ij}^-} \tag{3}$$

s.t. $\sum_{E_{ij} \in E} P(E_{ij} | \omega, \theta) = 1$.

Note that $P(E_{ij} | \omega, \theta) = P(E_{ji} | \omega, \theta)$ since there are no difference between the edges $E_{ij}$ and $E_{ji}$. Finally, the marginal likelihood of the signed network can be written as

$$P(A | \omega, \theta) = P(E | \omega, \theta) = \prod_{E_{ij} \in E} (\sum_{rr} \omega_{rr} \theta_{ri} \theta_{rj})^{E_{ij}^+} \cdot (\sum_{rs(r \neq s)} \omega_{rs} \theta_{ri} \theta_{sj})^{E_{ij}^-} \tag{4}$$

The parameters of Eq. (4) cannot be estimated using the likelihood maximization estimation because the chosen community pair $\{r', s'\}$ of an edge is a hidden variable. In our study, we use the EM algorithm for parameter estimation, which is a general approach to estimate the parameters of probabilistic mixture model by maximizing the expected likelihood iteratively. In each iteration, it computes the posterior probabilities of hidden variables using model parameters in E step, and re-estimates the model parameters in M step.

The log-likelihood function of Eq. (4) is

$$L = \ln P(A | \omega, \theta) = \ln P(E | \omega, \theta)$$
$$= \sum_{E_{ij} \in E} \left\{ E_{ij}^+ \ln(\sum_{rr} \omega_{rr} \theta_{ri} \theta_{rj}) + E_{ij}^- \ln(\sum_{rs(r \neq s)} \omega_{rs} \theta_{ri} \theta_{sj}) \right\} \tag{5}$$

It is usually converted to an expected log-likelihood function as Eq.(6) using the Jensen's inequality because it is difficult to be optimized directly.

$$L = \sum_{E_{ij} \in E} \left\{ E_{ij}^+ \ln(\sum_{rr} \omega_{rr} \theta_{ri} \theta_{rj}) + E_{ij}^- \ln(\sum_{rs(r \neq s)} \omega_{rs} \theta_{ri} \theta_{sj}) \right\}$$

$$\geq \sum_{rr} P(r, r \mid E^+, \omega, \theta) \ln P(E^+ \mid r, \omega, \theta)$$

$$+ \sum_{rs(r \neq s)} P(r, s \mid E^-, \omega, \theta) \ln P(E^- \mid r, s, \omega, \theta) \quad , \quad (6)$$

$$= \sum_{E_{ij} \in E^+ rr} q_{ijrr} E_{ij}^+ (\ln \omega_{rr} + \ln \theta_{ri} + \ln \theta_{rj})$$

$$+ \sum_{E_{ij} \in E^- rs(r \neq s)} Q_{ijrs} E_{ij}^- (\ln \omega_{rs} + \ln \theta_{ri} + \ln \theta_{sj})$$

where $q_{ijrr} = P(r, r \mid E^+, \omega, \theta)$ and $Q_{ijrs(r \neq s)} = P(r, s \mid E^-, \omega, \theta)$ respectively denote the probabilities of a positive link from a community $r$ and a negative link from different communities $r$ and $s$.

In the E step, the algorithm calculates the posterior probability of hidden variable ($r'$, $s'$) (i.e., $q_{ijrr}$ and $Q_{ijrs}$) using $\omega$ and $\theta$. They can be calculated by

$$q_{ijrr} = P(r \mid E^+, \omega, \theta)$$

$$= \frac{P(r, E^+ \mid \omega, \theta)}{P(E^+ \mid \omega, \theta)} \quad (7)$$

$$= \frac{\omega_{rr} \theta_{ri} \theta_{rj}}{\sum_{rr} \omega_{rr} \theta_{ri} \theta_{rj}}$$

$$Q_{ijrs(r \neq s)} = P(r, s \mid E^-, \omega, \theta)$$

$$= \frac{P(r, s, E^- \mid \omega, \theta)}{P(E^- \mid \omega, \theta)} \quad (8)$$

$$= \frac{\omega_{rs} \theta_{ri} \theta_{sj}}{\sum_{rs(r \neq s)} \omega_{rs} \theta_{ri} \theta_{sj}}$$

In the M step, the algorithm re-estimates $\omega$ and $\theta$ using $q_{ijrr}$ and $Q_{ijrs}$ from the E step. To estimate $\omega$ and $\theta$, we optimizes the expected log-likelihood function in Eq. (6). Considering that $\sum_{rr} \omega_{rr} + \sum_{rs(r \neq s)} \omega_{rs} = 1$ and $\sum_{i} \theta_{ri} = 1$, we obtain the Lagrange form of Eq. (6) as follows:

$$\tilde{L} = \sum_{E_{ij} \in E^+ rr} q_{ijrr} E_{ij}^+ (\ln \omega_{rr} + \ln \theta_{ri} + \ln \theta_{rj})$$

$$+ \sum_{E_{ij} \in E^- rs(r \neq s)} Q_{ijrs} E_{ij}^- (\ln \omega_{rs} + \ln \theta_{ri} + \ln \theta_{sj})$$

$$+ \rho(1 - \sum_{rr} \omega_{rr} - \sum_{rs(r \neq s)} \omega_{rs})$$

$$+ \sum_{rr} \gamma_r (1 - \sum_i \theta_{ri})$$

(9)

where $\rho$, $\gamma_r$ are the Lagrange multipliers. All parameters are derived by setting the derivative of $\tilde{L}$ to be 0:

$$\begin{cases} \omega_{rr} = \dfrac{\sum\limits_{E_{ij} \in E^+} q_{ijrr} E_{ij}^+}{\sum\limits_{E_{ij} \in E^+ rr} q_{ijrr} E_{ij}^+ + \sum\limits_{E_{ij} \in E^- rs(r \neq s)} Q_{ijrs} E_{ij}^-} \\ \omega_{rs(r \neq s)} = \dfrac{\sum\limits_{E_{ij} \in E^-} Q_{ijrs} E_{ij}^-}{\sum\limits_{E_{ij} \in E^+ rr} q_{ijrr} E_{ij}^+ + \sum\limits_{E_{ij} \in E^- rs(r \neq s)} Q_{ijrs} E_{ij}^-} \\ \theta_{ri} = \dfrac{\sum\limits_{j} q_{ijrr} E_{ij}^+ + \sum\limits_{js(r \neq s)} Q_{ijrs} E_{ij}^-}{\sum\limits_{E_{ij} \in E^+} q_{ijrr} E_{ij}^+ + \sum\limits_{E_{ij} \in E^- s(r \neq s)} Q_{ijrs} E_{ij}^-} \end{cases}$$

(10)

Once the model parameters are estimated as in Eq. (10), the probability of node $i$ belonging to community $r$ denoted by $\alpha_{ir}$, can be calculated by

$$\alpha_{ir} = \frac{\sum\limits_s \omega_{rs} \theta_{ri}}{\sum\limits_{rs} \omega_{rs} \theta_{ri}}$$

(11)

It means that a node can belong to several communities simultaneously. Therefore, the proposed model provides a soft-partition of the network with soft memberships of nodes, not a hard-partition. If we want to get a hard-partition, we can simply assign each node $i$ to the community it most likely belongs to. That is $r = \arg\max\{\alpha_{1r}, \alpha_{2r}, ..., \alpha_{Kr}\}$.

Suppose that the number of $E^+$ is $l^+$ and the number of $E^-$ is $l^-$, the time complexity of calculating $q_{ijr}$ and $Q_{ijrs(r \neq s)}$ in the E step are $O(l^+ \times K)$ and $O(l^- \times K(K-1))$ respectively. Thus, the total time cost of E step is $O(l^+ \times K + l^- \times K(K-1))$. In the M step, we need to calculate $\omega_{rr}$, $\omega_{rs(r \neq s)}$ and $\theta_{ri}$, the corresponding time-complexities are $O(l^+ \times K + l^- \times K(K-1))$, $O(l^+ \times K + l^- \times K(K-1))$ and $O(l^+ + l^- \times K)$. Then, the total time cost of M step is $O(l^+ \times K + l^- \times K^2)$. If the EM algorithm converges within $T$ iterations, the time-complexity of the SPM model will be $O(T(l^+ \times K + l^- \times K^2))$.

### 3. Two extreme signed networks

Particularly, in a signed network with only positive links, that is $E^+ = E$, our model is simplified to

$$P(A|\omega,\theta) = P(E|\omega,\theta) = \prod_{E_{ij} \in E}(\sum_{rr}\omega_{rr}\theta_{ri}\theta_{rj})^{E_{ij}^+} \tag{12}$$

In the E step,

$$q_{ijrr} = \frac{\omega_{rr}\theta_{ri}\theta_{rj}}{\sum_{rr}\omega_{rr}\theta_{ri}\theta_{rj}} \tag{13}$$

In the M step,

$$\begin{cases} \omega_{rr} = \dfrac{\sum_{E_{ij} \in E^+} q_{ijrr} E_{ij}^+}{\sum_{E_{ij} \in E^+ rr} q_{ijrr} E_{ij}^+} \\ \theta_{ri} = \dfrac{\sum_{j} q_{ijrr} E_{ij}^+}{\sum_{E_{ij} \in E^+} q_{ijrr} E_{ij}^+} \end{cases} \tag{14}$$

Our algorithm is similar to the Simple Probabilistic Algorithm Expectation Maximization (SPEAM) model [29], for assortative structure detection, where all edges are in communities.

Similarly, in another signed network with only negative links, that is $E^- = E$, our model is simplified to

$$P(A|\omega,\theta) = P(E|\omega,\theta) = \prod_{E_{ij} \in E}(\sum_{rs(r \neq s)}\omega_{rs}\theta_{ri}\theta_{sj})^{E_{ij}^-} \tag{15}$$

In the E step,

$$Q_{ijrs(r \neq s)} = \frac{\omega_{rs}\theta_{ri}\theta_{sj}}{\sum_{rs(r \neq s)}\omega_{rs}\theta_{ri}\theta_{sj}} \tag{16}$$

In the M step,

$$\begin{cases} \omega_{rs(r \neq s)} = \dfrac{\sum_{E_{ij} \in E^-} Q_{ijrs} E_{ij}^-}{\sum_{E_{ij} \in E^- rs(r \neq s)} Q_{ijrs} E_{ij}^-} \\ \theta_{ri} = \dfrac{\sum_{js(r \neq s)} Q_{ijrs} E_{ij}^-}{\sum_{E_{ij} \in E^- s(r \neq s)} Q_{ijrs} E_{ij}^-} \end{cases} \tag{17}$$

Our algorithm can be used to identify the disassortative structure, where all edges are across communities.

## 4. Experiment and analysis

To investigate the effectiveness of the SPM model on overlapping community detection in signed networks. We first test it on a large number of signed networks including a signed network with only positive links, a signed network with only negative links, an illustrative network, two real-world networks and a series of synthetic networks. Then we discuss the model selection issue—how to determine the optimal number of communities.

*4.1   Community detection in a signed network with only positive links*

The Zachary club network, which characterizes the acquaintance relationship between 34 members [36], is used to test the capability of our model on assortative structure detection in a signed network with only positive links. The club network is split into two groups because of a dispute between the administrator and karate teacher. It has been used as a common dataset for overlapping community detection in many studies. Figure 2 shows the communities detected by our algorithm when setting $K$=2. The SPM model correctly identifies two assortative structures with several overlapping nodes: {3, 9, 14, 20, 31, 32}. To further investigate the effectiveness of our model, we compared it with several popular models, including Generalized Stochastic Blockmodel (GSB) [25], Newman Mixture Model (NMM) [9] and SPEAM [29]. Table 1 shows the memberships of the 6 overlapping nodes when using different models. The numbers in a parenthesis are coefficients indicating how strongly a node belongs to all communities (called community coefficients). For example, the first (0.51, 0.49) in the first row indicates that node 3 belongs to two communities with probabilities of 0.51 and 0.49 respectively. We can see that our model gets the same result as GSB and SPAEM, and a better result than NMM. It is easy to understand that the results from GSB, SPAEM and our model are the same because a signed network with only positive links only contains assortative structures. Each assortative structure is a community. It is not surprised that our model outperforms NMM since GSB has been proved superior to NMM on the Zachary club network in [25].

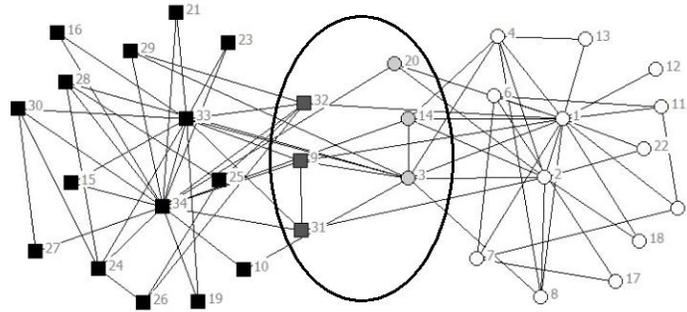

**Figure 2.**  The network of the Zachary club with 34 nodes and 78 positive links. The real communities in this network are marked by different type of shapes: square and circle. The shading nodes in the ellipse are overlapping nodes of soft memberships identified by our algorithm.

**Table 1.**  The memberships of 6 overlapping nodes when using different models.

| NodeID | GSB | NMM | SPEAM | SPM |
|---|---|---|---|---|
| 3 | (0.51, 0.49) | (1.00, 0.00) | (0.51, 0.49) | **(0.51, 0.49)** |
| 9 | (0.30, 0.70) | (0.04, 0.96) | (0.30, 0.70) | **(0.30, 0.70)** |
| 14 | (0.76, 0.24) | (1.00, 0.00) | (0.76, 0.24) | **(0.76, 0.24)** |
| 20 | (0.67, 0.33) | (0.87, 0.13) | (0.67, 0.33) | **(0.67, 0.33)** |
| 31 | (0.29, 0.71) | (0.08, 0.92) | (0.29, 0.71) | **(0.29, 0.71)** |
| 32 | (0.17, 0.83) | (0.00, 1.00) | (0.17, 0.83) | **(0.17, 0.83)** |

*4.2   Community detection in a signed network with only negative links*

We adopt the dataset used in [37] to test the capability of our model on disassortative structure detection in a signed network with only negative links. The dataset is a network of 112 common adjectives and nouns in the novel David Copperfield by Charles Dickens connected by 425 edges. Each edge in the network denotes a pair of adjacent words in the text. To test our model on the dataset, we change the original edges into negative links. Figure 3 shows the communities detected by our model when setting $K$=2. Our model detects a bipartite structure, which is composed of two disassortative structures: an adjective group and a noun group. In addition, we also compare our model with GSB, NMM and SPAEM. Their performance is measured by the node accuracy of the hard-partition derived from them. 100 of the 112 nodes are correctly classified by GSB, NMM and our model, while only 60 of the 112 nodes are correctly classified by SPAEM. It means that SPAEM is worse than GSB, NMM and our model on dissortative structure detection in signed networks with only negative links. The reason is that the SPAEM assumes that networks are composed of assoratative structures.

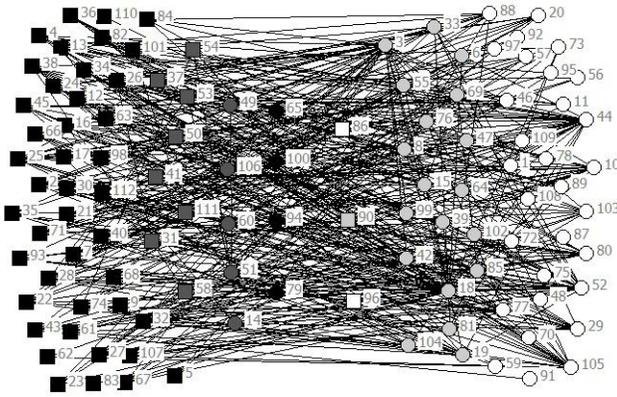

**Figure 3.** The network of 112 common adjectives and nouns in the novel David Copperfield by Charles Dickens connected by 425 negative links. The adjective and noun groups (i.e., communities) are denoted by circles and squares, respectively. The shading nodes are overlapping nodes of soft memberships identified by our algorithm.

*4.3 Overlapping community detection in an illustrative signed network*

We test our model on the illustrative signed network shown in figure 1. It contains 9 nodes connected by 16 positive links and 9 negative links. The nodes fall into two overlapping communities with two overlapping nodes (i.e., E and F). When setting the number of communities $K = 2$, our model correctly detect two communities with two overlapping nodes as shown in figure 4, where the numbers are community coefficients of nodes. It is very clear that nodes (A, B, C, D) completely belong to the left community since their community coefficients are (1.0, 0.0), nodes (G, H, I) completely belong to the right community since their community coefficients are (0.0, 1.0), and nodes (E, F) belong to the two communities simultaneously since their community coefficients are (0.43, 0.57) and (0.34, 0.66) respectively.

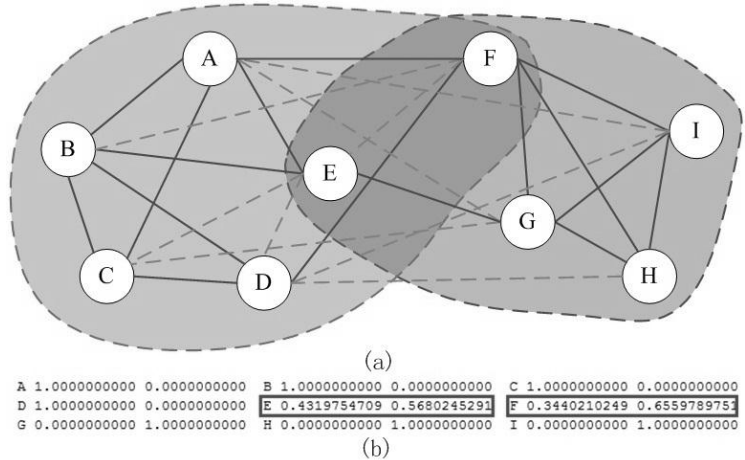

**Figure 4.** Overlapping community detection in the illustrative signed network shown in figure 1. (a) The overlapping communities are detected by the SPM model. (b) The community coefficients of all nodes predicted by the SPM model.

*4.4 Overlapping community detection in real-world signed networks*

We test our model on two public datasets, which are widely used for community detection.

The first signed network is a relation network of 10 parties of the Slovene Parliamentary in 1994 [38] as shown in figure 5(a). The numbers are the weights of links in the network estimated by 72 questionnaires among 90 members of Slovene National Parliament. The questionnaires were designed to estimate the distance of 10 parties on the scale from -3 to 3, and the final weight were the averaged value multiplying by 100. The 10 parties fall into two communities: (1, 3, 6, 8, 9) and (2, 4, 5, 7, 10), a hard-partition of the network. When setting $K=2$, our model detected two communities with an overlapping node as shown in figure 5(b). The community coefficients of nodes are shown in figure 5(c). The overlapping communities detected by our model are (1, 3, 6, 8, 9, 10) and (2,

4, 5, 7, 10), which are a little different with the real communities. The difference is reasonable because this network is designed for finding a hard-partition, not a soft-partition. For node 10, it really does not completely belong to the community on the right because there are two positive links (10-2, 10-4) and two negative links (10-5, 10-7) related to it in that community. On the other hand, if we use the method mentioned in section 2 to convert the soft-partition predicted by our model into a hard-partition, the hard-partition will be the same as the real one.

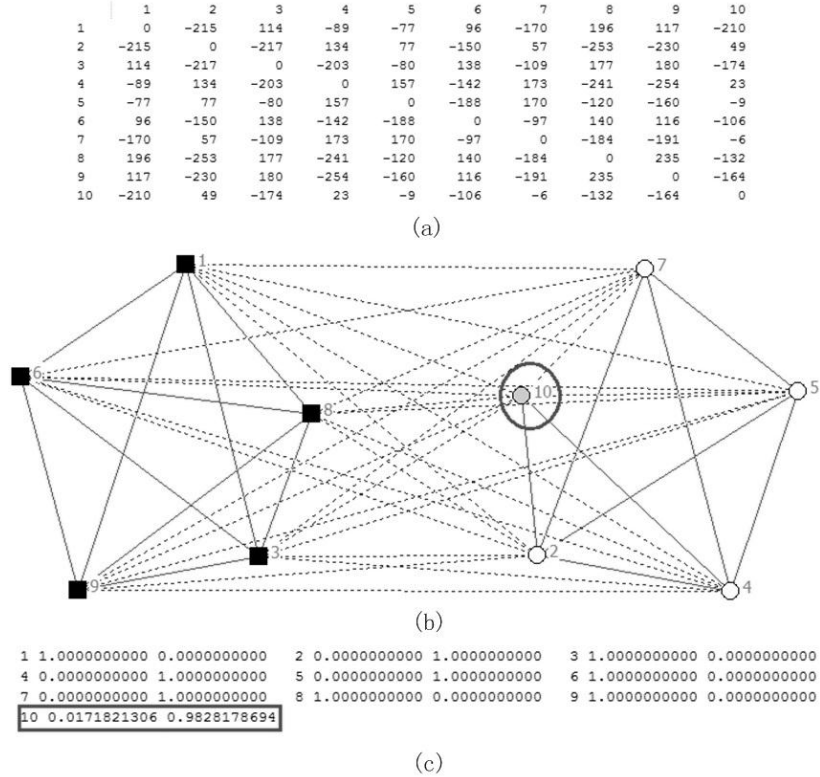

**Figure 5.** Overlapping community detection in the Slovene Parliamentary signed network. (a) The adjacency matrix of the relation network of 10 Slovene Parliamentary parties. (b) The overlapping communities are detected by the SPM model. The solid lines denote the positive links, and the dotted lines denote the negative links. The real communities are marked by different types of shapes (i.e., square and circle). The shading nodes are nodes of soft memberships identified by our model. (c) The community coefficients of all nodes obtained by the SPM model. All overlapping nodes are emphasized in (b) and (c).

The second signed network is the Gahuku-Gama Subtribes network about the cultures of highland New Guinea [39] as shown in figure 6(a). It describes the political alliance and enmities among the 16 Gahuku-Gama subtribes. The positive and negative links of the network correspond to different political arrangements. The 16 subtribes fall into three communities. Among them, one subtribe sides with two communities. When we apply our model on this network with $K=3$, three communities are correctly detected with an overlapping node as shown in figure 6(b). The community coefficients of nodes are shown in figure 6(c).

*4.5   Community detection in synthetic signed networks*

It is common to validate the performance of algorithms for community detection on synthetic networks. In our study, we also test the SPM model on some synthetic signed networks. The synthetic signed networks are generated using the method proposed by Yang [17]. We use $SG(c, (n_1,n_2,...,n_c), k, p_{in}, p_+, p_-)$ to denote a synthetic signed network, where $c$ is the number of communities, $(n_1,n_2,...,n_c)$ are the number of nodes of each community, $k$ is the degree of each node, $p_{in}$ is the probability of each node connecting with other nodes in the same community, $p_+$ denotes the probability of positive links across communities and $p_-$ denotes the probability of negative links within communities. Note that we simplify it as $SG(c, n, k, p_{in}, p_+, p_-)$ when $n_1=n_2=...=n_c$. We test our model on two types of synthetic signed networks: partitionable signed networks in which both $p_+$ and $p_-$ are 0; non-partitionable signed networks in which $p_+$ or $p_-$ is not 0. In addition, we conduct a number of experiments to test the robustness of the SPM model.

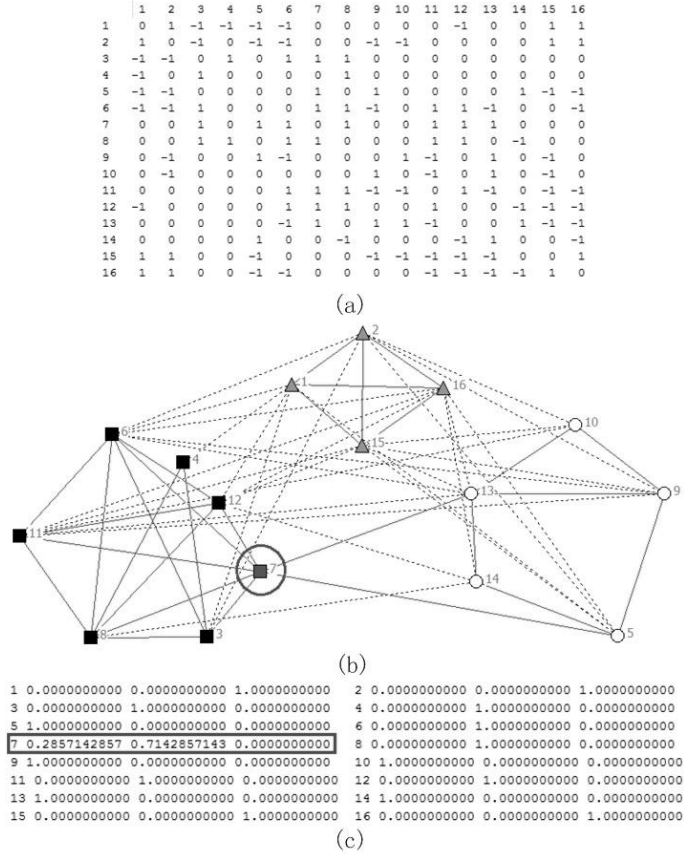

**Figure 6.** Overlapping community detection in the Gahuku-Gama Subtribes signed network. (a) The adjacency matrix of a network of 16 Gahuku-Gama Subtribes. (b) The overlapping communities are detected by the SPM model. The solid lines denote the positive links, and the dotted lines denote the negative links. The real communities are marked by different types of shapes: square, circle and triangle. The shading nodes are overlapping nodes identified by the SPM model. (c) The community coefficients of all nodes obtained by the SPM model. All overlapping nodes are emphasized in (b) and (c).

The performance of all models is measured by the Normalized Mutual Information (NMI) [40], which is a widely used method for evaluating the community detection:

$$P_{nmi}(G,G') = \frac{2MI(G,G')}{H(G)+H(G')}, \qquad (18)$$

where $G=(G_1,G_2,...,G_K)$ are defined communities, $G'=(G'_1,G'_2,...,G'_K)$ are communities detected by an algorithm, $H(G)$ and $H(G')$ are the entropies of $G$ and $G'$, and $MI(G,G')$ is the mutual information between them. A high $P_{nmi}$ means a good detection. Specially, $P_{nmi}=1$ means that the detection is perfect.

**Partitionable signed networks**

We test our model on three partitionable synthetic signed networks: $SG(4, 30, 16, 0.8, 0, 0)$ as shown in figure 7(a), $SG(4, 30, 16, 0.1, 0, 0)$ as shown in figure 7(c) and $SG(20, 30, 16, 0.8, 0, 0)$ as shown in figure 7(e). The positive links are denoted by white points; the negative links are denoted by black points in the figures. The first network has the same parameters as the second one except the density of edges in each community. The first network has the same parameters as the third one except the number of communities. The three networks are used to test the effect of density of edges in a community on our model, and the effect of the number of communities on our model. When our model is applied on these three networks, all communities are correctly detected as shown in figure 7(b), 7(d) and 7(f) respectively. The $P_{nmi}$ of our model on all three networks is 1. The results show that the SPM model is unaffected by not only the density of edges in each community ($SG(4, 30, 16, 0.8, 0, 0)$ vs $SG(20, 30, 16, 0.1, 0, 0)$), but also the number of communities ($SG(4, 30, 16, 0.8, 0, 0)$ vs $SG(4\ 30, 16, 0.8, 0, 0)$).

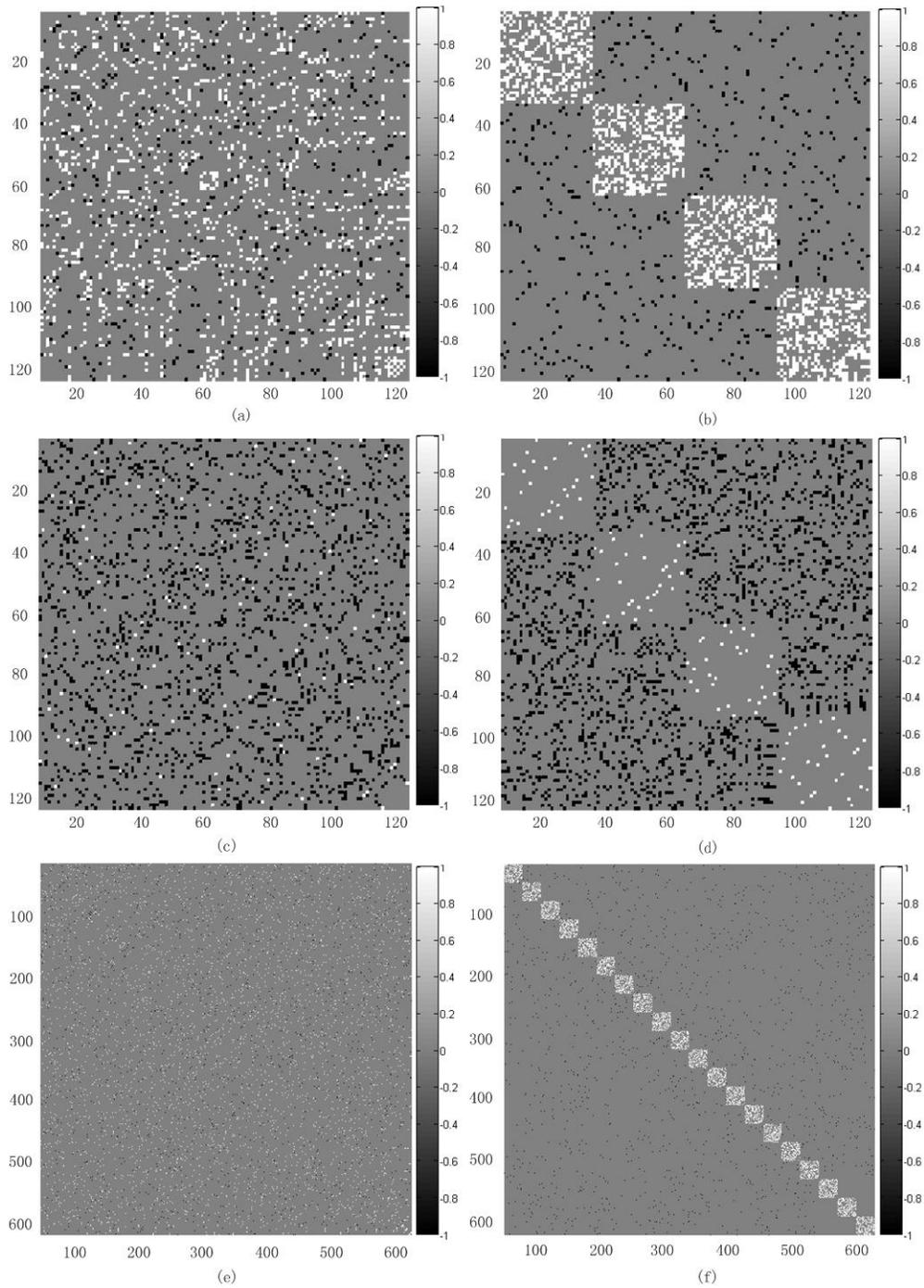

**Figure 7.** Community detection on three partitionable synthetic signed networks: $SG(4, 30, 16, 0.8, 0, 0)$, $SG(4, 30, 16, 0.1, 0, 0)$ and $SG(20, 30, 16, 0.8, 0, 0)$. (a) The adjacency matrix of the first network. (b) The communities detected by the SPM model on the first network. (c) The adjacency matrix of the second network. (d) The communities detected by the SPM model on the second network. (e) The adjacency matrix of the third network. (f) The communities detected by the SPM model on the third network.

**Non-partitionable signed networks**

We also test our model on two non-partitionable synthetic signed networks: $SG(4, 30, 16, 0.8, 0.2, 0.2)$ as shown in figure 8(a) and $SG(4, (30, 60, 90, 120), 16, 0.8, 0.2, 0.2)$ as shown in figure 8(c). All parameters of the first network are the same as the first network in figure 7(a) except $p_+$ and $p_-$. In order to test the effect of noises on our model, we set both $p_+$ and $p_-$ to 0.2. The second network has the same parameters as the first one except the number of edges in each community. In the first network, all communities compose of the same number of nodes. In the second network, the number of nodes in a community is different with each other's. The two networks are used to test the effect of noise on our model, and the effect of the number of edges in each community. When our model is applied on these two networks, all communities are correctly detected as shown in figure 8(b) and figure 8(d) respectively. The $P_{nmi}$ of our model on both two networks is 1. The results show that the SPM model is unaffected when either adding noises ($SG(4, 30, 16, 0.8, 0.2, 0.2)$ vs $SG(4, 30, 16, 0.8, 0, 0)$) or setting different numbers of nodes in communities ($SG(4, 30, 16, 0.8, 0.2, 0.2)$ vs $SG(4, (30, 60, 90, 120), 16, 0.8, 0.2, 0.2)$).

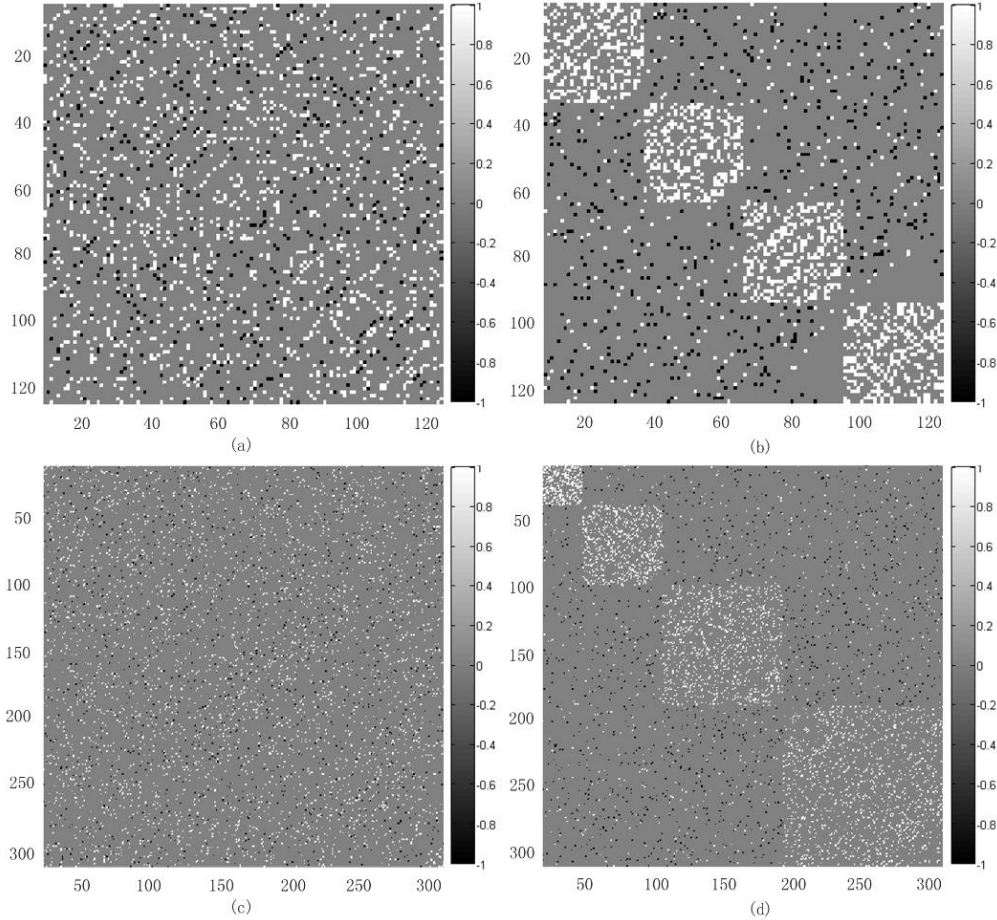

**Figure 8.** Community detection on two non-partitionable synthetic signed networks: $SG(4, 30, 16, 0.8, 0.2, 0.2)$ and $SG(4, (30, 60, 90, 120), 16, 0.8, 0.2, 0.2)$. (a) The adjacency matrix of the first network. (b) The communities detected by the SPM model on the first network. (c) The adjacency matrix of the second network. (d) The communities detected by the SPM model on the second network.

**Robust evaluation**

We not only compare the SPM model with the Signed Newman Mixture (SNM) model [20], which needs a pre-defined number of communities, but also compare it with the FEC model [17] and Traag's model [19], which do not need a pre-defined number of communities. For comparison, we construct two types of signed works: $SG(4, 30, 16, p_{in}, 0, 0)$ with $p_{in}$ gradually changing from 0 to 1, and $SG(4, 30, 16, 0.8, p_+, p_-)$ with both $p_+$ and $p_-$ gradually changing from 0 to 1. The results on them are shown in figure 9 and figure 10. Note that results at some points are not displayed in figure 9 and figure 10 as there is no ground-truth community at them. For example, when $p_{in}= 0.0$, there is no ground-truth community in $SG(4, 30, 16, p_{in}, 0, 0)$ as there is no positive link in any community. When $p_+>0.5$, there is also no ground-truth community in $SG(4, 30, 16, 0.8, p_+, p_-)$ as the positive links in any community is less than the positive links across communities. When $p_->0.5$, there is no ground-truth community in $SG(4, 30, 16, 0.8, p_+, p_-)$ too as the negative links across communities is less than the negative links in any community.

In figure 9, each curve is the average $P_{nmi}$ of a model with $p_{in}$ on 30 synthetic random networks. All the models are applied on the same networks. The $P_{nmi}$ of the SPM model is always 1 when $0.1 \leq p_{in} < 1$. When $0.05 \leq p_{in} < 0.15$, the SPM model slightly outperforms the SNM model and Traag's model, significantly outperformed the FEC model. When $p_{in} \geq 0.15$, the $P_{nmi}$ of the SNM model, and Traag's model achieves 1, which is the same as the SPM model. For the FEC model, the $P_{nmi}$ achieves 1 when $p_{in} \geq 0.6$. Overall, the SPM model outperforms the other three models when $p_{in}$ gradually changes from 0.05 to 1.

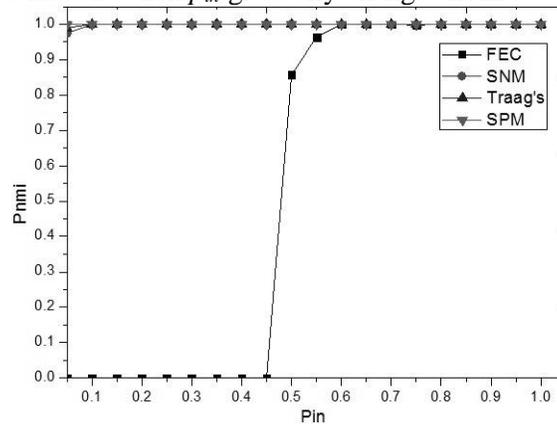

**Figure 9.** Community detection performance of four algorithms on the networks SG(4, 30, 16, $p_{in}$, 0, 0).

In figure 10, a surface is also the average $P_{nmi}$ of a model with $p_+$ and $p_-$ on 30 synthetic random networks. The $P_{nmi}$ of the SPM model is much higher than the SNM model when $0 \leq p_+ \leq 0.5, 0 \leq p_- \leq 0.5$. Compared with the FEC model, the SPM model achieves the same result $P_{nmi}=1$ when $0 \leq p_+ \leq 0.3, 0 \leq p_- \leq 0.5$, and achieves higher $P_{nmi}$ when $0.3 < p_+ \leq 0.5, 0 \leq p_- \leq 0.5$. Compared with the Traag's model, the SPM model achieves the similar result when $0 \leq p_+ \leq 0.45, 0 \leq p_- \leq 0.5$, and achieves much higher $P_{nmi}$ when $0.45 < p_+ \leq 0.5, 0.2 \leq p_- \leq 0.5$. Although the SPM model is slightly inferior to the Traag's model when $0.45 < p_+ \leq 0.5, 0 \leq p_- < 0.2$, the $P_{nmi}$ of the SPM model is still not less than 0.6, which is acceptable. Overall, the SPM model is superior to the SNM and FEC models, competitive with the Traag's model.

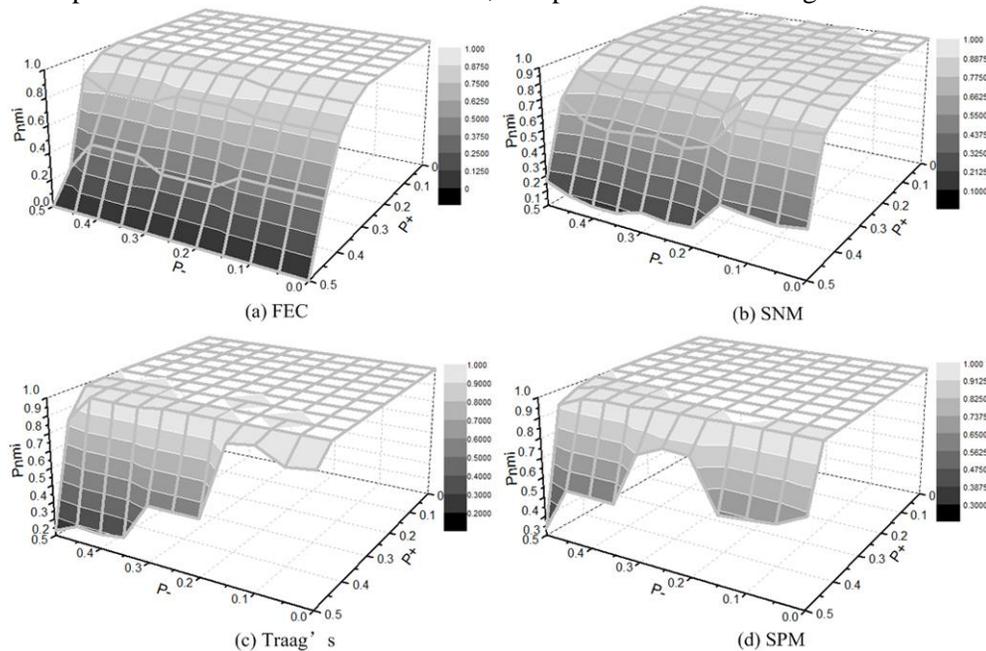

**Figure 10.** Community detection performance of four algorithms on the network $SG(4, 30, 16, 0.8, p_+, p_-)$.

In summary, experiments on networks with only positive links or negative links show that the SPM model correctly identifies assortative structures or disassortative structures as the same as other state-of-the-art models; experiments on real-world signed networks show that the SPM model is able to detect overlapping communities which are neglected by most of the current popular models; experiments on synthetic signed networks show that the SPM model outperforms other state-of-the-art models at shedding light on community detection in signed networks.

*4.6    Model selection issue*

A limit of our model is that it requires a predefined community number, which is usually unknown or uncertain in many real-world networks. Therefore, it is necessary to provide a criterion to determine the community number for our model. We test our model under two criteria: the minimum description length (MDL) principle [41], and the criterion in [42].

The MDL is a popular criterion for model selection issue which contains two parts: one describing the coding length of the networks; the other one describing the length of model parameters. For our model, the coding length is $-L/2$, and the length of model parameters is $-\sum_{rs} \omega_{rs} \ln \omega_{rs} - \sum_{ri} \theta_{ri} \ln \theta_{ri}$. We apply the MDL to our model on the aforementioned signed networks. The MDL fails to acquire the number of communities in all networks. When the MDL is applied to the SNM model, it fails too. It seems that the MDL is not suitable for our model as well as the SNM model.

The error criterion function presented in [42] can be written as
$$P(C) = \eta N + (1-\eta)P, \tag{19}$$
where $N$ denotes the total number of negative links within communities, $P$ denotes the total number of positive links between communities, and $\eta$ denotes the weight of negative links ($0 \leq \eta \leq 1$). It can be used to determine the community number of signed networks because there exists only one partition to make the criterion function (Eq.19) minimum for any signed network according to the theorem in [43]. We used the error criterion function to determine the community number on the signed networks aforementioned for the SPM model. The results are shown in figure 11. The SPM model finds one optimal community number on the Slovene Parliamentary Party network, the Gahuku-Gama Subtribes network, the $SG$(4, 30, 16, 0.1, 0, 0) network and the $SG$(4, (30, 60, 90, 120), 16, 0.8, 0.2, 0.2) network, which are correct. On the illustrative network, the $SG$(4, 30, 16, 0.8, 0, 0) network, the $SG$(20, 30, 16, 0.8, 0, 0) network and the $SG$(4, 30, 16, 0.8, 0.2, 0.2) network, the SPM model finds multiple optimal community numbers. By checking community coefficients of nodes in each network of them, we find that all the optimal community numbers correspond to the same hard-partition, which is also correct. Thus, the error criterion function is suitable to determine the community number for the SPM model.

**5.    Conclusions**

In this paper, we proposed a novel probabilistic model for overlapping community detection in signed networks. The proposed model is a variant of the probabilistic mixture model. The advantages of the model are (i) providing soft-partition solutions for signed networks; (ii) providing soft-memberships of nodes. Experiments on a number of real-world and synthetic signed networks show that our SPM model: (i) can identify assortative structures or disassortative structures as the same as other state-of-the-art models; (ii) can detect overlapping communities; (iii) outperform other state-of-the-art models at shedding light on the community detection in synthetic signed networks. In addition, the general criterion function is proved suitable to determine the optimal number of communities. As future work, we will apply our model to community detection on real scalable signed networks, and seek possible applications.

**Acknowledgements**
The work is partially supported by the National Natural Science Foundation of China (61272383), the China Postdoctoral Science Foundation (2011M500669) and the Strategic Emerging Industry Development Special Fund of Shenzhen (ZDSY20120613125401420). We also thank all researchers who release the data sets used in this study.

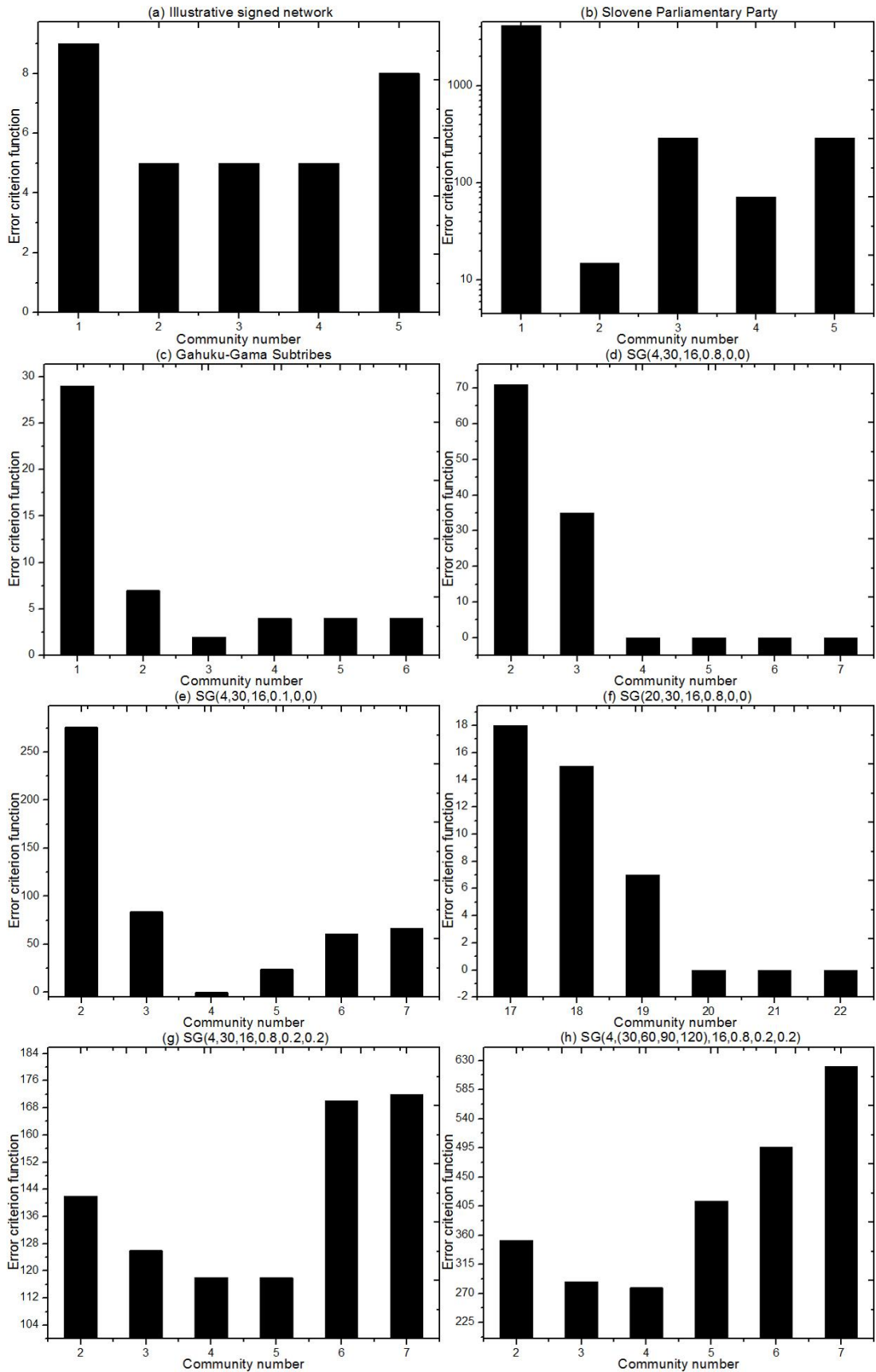

**Figure 11.** The optimal community number(s) determined by the error criterion function presented in [42].